\documentclass[12pt]{article}
\usepackage{aaspp4}

\newcommand \lovdl{\relax \ifmmode \lambda/\Delta\lambda \else $\lambda/\Delta\lambda$\fi}
\newcommand \euve{{\it EUVE}}
\newcommand \orfeus{{\it ORFEUS-SPAS II}}
\newcommand \iue{{\it IUE}}

\newcommand \copernicus{{\it Copernicus}}
\newcommand \ecma{\hbox{$\epsilon$ CMa}}
\newcommand \bcma{\hbox{$\beta$ CMa}}
\newcommand \etal{et~al.}
\newcommand \Teff{\relax \ifmmode {T_{\rm eff}} \else $T_{\rm eff}$ \fi}
\newcommand \mm{\relax \ifmmode {\, \mu{\rm m}} \else $\mu{\rm m}$\fi}

\begin{document}

\title{{\it ORFEUS-SPAS II} EUV Spectroscopy of \ecma\ (B2 II)\footnotemark}

\author{David H. Cohen}

\affil{Fusion Technology Institute and Department of Astronomy, \\
  University of Wisconsin--Madison, 1500 Engineering Drive, Madison,
  WI 53706 \\ e-mail: cohen@duff.astro.wisc.edu}

\author{Mark Hurwitz}

\affil{Space Sciences Laboratory, University of California, Berkeley,
CA 94720 \\ e-mail: markh@ssl.berkeley.edu}

\author{Joseph P. Cassinelli}

\affil{Department of Astronomy, University of Wisconsin--Madison, \\
475 Charter St., Madison, WI 53706 \\ e-mail:
cassinelli@madraf.astro.wisc.edu}

\author{Stuart Bowyer}

\affil{Space Sciences Laboratory, University of California, Berkeley,
CA 94720 \\ e-mail: bowyer@ssl.berkeley.edu}

   \footnotetext{Based on the development and utilization of {\it
   ORFEUS} ({\it Orbiting and Retrievable Far and Extreme Ultraviolet
   Spectrometers}), a collaboration of the Institute for Astronomy and
   Astrophysics at the University of T\"{u}bingen, the Space
   Astrophysics Group of the University of California at Berkeley, and
   the Landessternwarte Heidelberg.}



\begin{abstract}

We report on extreme-ultraviolet (EUV) spectroscopic observations of
the B bright giant $\epsilon$ Canis Majoris made during the \orfeus\/
mission.  We assess the performance of the instrument in the EUV and
find that the effective area is roughly 3 times that of the {\it
Extreme-Ultraviolet Explorer} (\euve) long-wavelength spectrometer and
that the spectral resolution is $\lovdl \approx 1250$.  We identify
most of the features, qualitatively compare different models, and
examine the wind-broadened \ion{O}{5} and \ion{Si}{4} lines, which
display blue edge velocities up to 800 km s$^{-1}$.

\end{abstract}


\section{Introduction}  

One of the surprising discoveries of the {\it Extreme-Ultrviolet
Explorer} (\euve) telescope was that the B star \ecma\ (B2 II) is the
brightest extrasolar source of radiation between 504 \AA\ and 760 \AA\
\markcite{vvw93}(Vallerga, Vedder, \& Welsh 1993).  This star is one
of only two early-type stars for which high quality
extreme-ultraviolet (EUV) spectra can be obtained.  As such, it
provides a rare opportunity to study the photosphere of a hot star
below the Lyman edge, where the continuum is formed very high in the
atmosphere and where wind effects and non-LTE effects are expected to
be more severe than in the UV or optical.  In this Letter, we report
on ORFEUS-Shuttle Pallet Satellite II (\orfeus) EUV observations of
\ecma\ between 520 and 665 \AA, made with unprecedented spectral
resolution.  We describe the performance of the \orfeus\/ Berkeley
spectrometer in the EUV, and show that the wealth of data provided in
the spectrum can be used to test wind models as well as photospheric
models of this early B star.

The very high EUV flux of \ecma\ is partially due to the extremely
small interstellar \ion{H}{1} column density on the sight-line to this
star ($N_H < 10^{18}$ cm$^{-2}$), but it is primarily caused by the
unexpectedly large intrinsic Lyman continuum flux.  In fact, the flux
levels observed by \euve\ are 30 times higher than the appropriate LTE
Kurucz \markcite{kur92}(1992) model atmosphere, and 100 times higher
than a non-LTE, line-blanketed TLUSTY \markcite{hl92}(Hubeny \& Lanz
1992) model \markcite{cass95}(Cassinelli \etal\ 1995).  A similar, but
not as extreme, excess is also seen in the \euve\ observation of the
B1 II-III star \bcma.  Both stars have mid-IR excesses in addition to
EUV excesses.  The IR continuum between 10 and 15 \mm\ in these B
stars is formed in the same physical layers as the Lyman continuum
between 500 and 700 \AA.  The IR continuum is dominated by a free-free
process, which is strictly thermal, and a local temperature excess of
about 2000 K over the Kurucz model fits both the Lyman and mid-IR
continua \markcite{cass95}(Cassinelli \etal\ 1995).

Different explanations of this purported temperature excess have been
put forward since the discovery of the large EUV fluxes in \ecma\/ and
\bcma.  These include additional line-blanketing and non-planar
effects \markcite{auf98}(Aufdenberg \etal\ 1998), mechanical heating
\markcite{cass96}(Cassinelli 1996), and X-ray irradiation from the
stellar wind \markcite{coh96}(Cohen \etal\ 1996).  An explanation for
the EUV excess that does not assume a temperature excess has also been
proposed.  It relies on non-LTE effects caused by Doppler shifts in
the subsonic regions of the outer atmosphere \markcite{naj96}(Najarro
\etal\ 1996).  Regardless of the mechanism, reproducing the
line-blanketed EUV spectrum would be an important constraint on any
model.

The nature of the winds of early B stars is largely unknown because of
the small number of wind-broadened lines visible in the optical and
UV.  Furthermore, without information about the ionization structure,
the interpretation of the limited number of wind lines seen in the UV
is ambiguous.  With the additional constraints imposed by the
measurement of EUV wind lines in \ecma, it should be possible to
determine the wind ionization balance and mass-loss rate of this early
B star.

The \euve\ spectroscopy, with an average resolution of $\lovdl \approx
300$, suffered from significant line-blending, making the measurement
of individual line strengths impossible.  Additionally, because of the
extreme line blanketing, it was not clear that the continuum level was
actually reached at any point between 504 \AA\ and the end of the
\euve\ sensitivity at 760 \AA.  With the new \orfeus\/ observations,
we can resolve single features in the \euve\ data into several
individual features. These higher resolution data will be very
valuable for putting tight constraints on wind models as well as
testing new atmosphere models.

\section{Data}

The data discussed in this Letter were collected with the Berkeley
spectrograph aboard the {\it ORFEUS-SPAS II} mission flown in 1996
November/December; \ecma\/ was observed five times, providing a
useful integration time of 9447 s.  The instrument is described in
Hurwitz \& Bowyer \markcite{hb86}\markcite{hb96}(1986, 1996) and the
performance of the \orfeus\/ mission is described in Hurwitz \etal\
\markcite{hur97} (1998).  To attenuate scattered far ultraviolet light
which would otherwise contaminate the EUV spectrum, \ecma\ was
observed through the tin-filtered aperture in diaphragm position 3.
This aperture is 120\arcsec\ in diameter, displaced off axis by
5\arcmin .0, and covered by a tin filter approximately 1500 \AA\
thick.  Extraction of the spectra and subtraction of background
generally followed the procedure discussed in Hurwitz \etal\
\markcite{hur97}(1998).

Characterization of the instrument performance in the extreme
ultraviolet is more difficult than in the far ultraviolet.  The only
detectable diffuse emission feature is He I 584.33 \AA.  We used about
one dozen relatively unblended features to define the overall
wavelength scale for these observations.  Residual errors in the
wavelength solution are below 0.5 \AA.  For the current analysis it is
not important to establish an absolute flux level, and we have not
attempted to do so in any detail here, although it is expected that
for filtered observations, the Berkeley spectrograph effective area is
about 3 times that of the \euve\/ spectrograph at wavelengths of
overlap. The peak transmission of the filter is only $\sim$ 20\%;
for white dwarfs and other sources that do not require the filter, the
effective area greatly exceeds that of \euve. 

\section{Discussion}

In order to identify lines, we compared the data with a TLUSTY $\Teff
= 21,000$ K, $\log g = 3.2$ non-LTE synthetic spectrum that uses the
Kurucz line-list (see \markcite{cass95}Cassinelli \etal\ 1995), as
well as with several hotter models ($\Teff = 23,000$ K and $\Teff =
24,700$ K).  The same lines are generally present in these different
models, although \ion{N}{2} and \ion{C}{2} are much stronger in the
nominal, 21,000 K, model and the balance between \ion{Fe}{4} and
\ion{Fe}{3} varies among the models.  We checked our identifications
against the line-list of Verner, Verner, \& Ferland
\markcite{ver96}(1996).  Overall, there is very good agreement between
these two references.  As can be seen in Figure \ref{fig:lineid}, many
strong lines and numerous weak ones were identified, with very few
features left unidentified.

\begin{figure}
\plotone{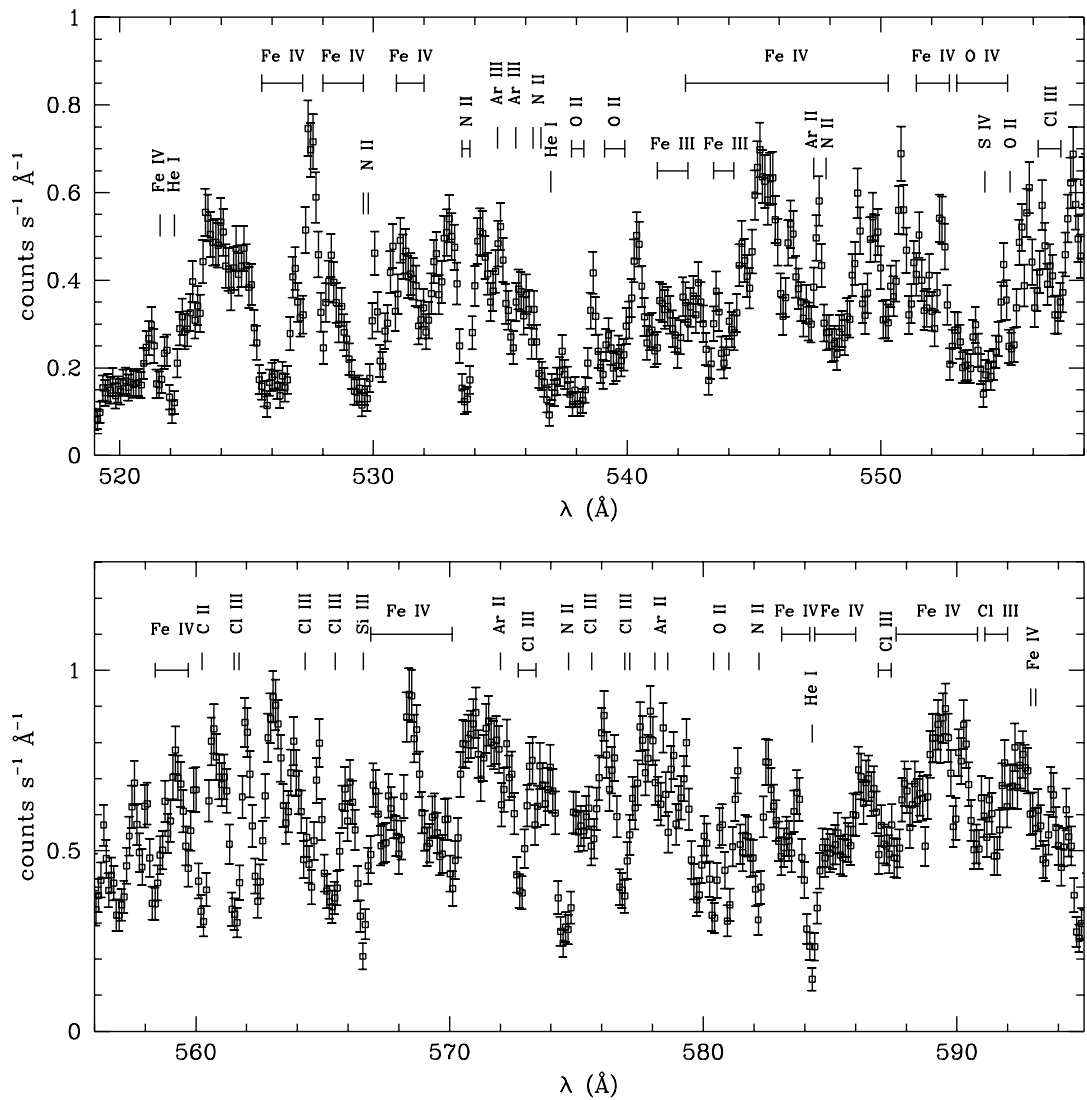}
\caption[]{Co-added \orfeus\/ spectrum from data collected over five
orbits, with bin sizes of 0.103 \AA\ and 1 $\sigma$ statistical
uncertainties indicated.  Strong lines are identified above the
spectrum.  Note that the wavelength scale is accurate to about 0.5
\AA.}
\label{fig:lineid}
\end{figure}

\setcounter{figure}{0}

\begin{figure}
\plotone{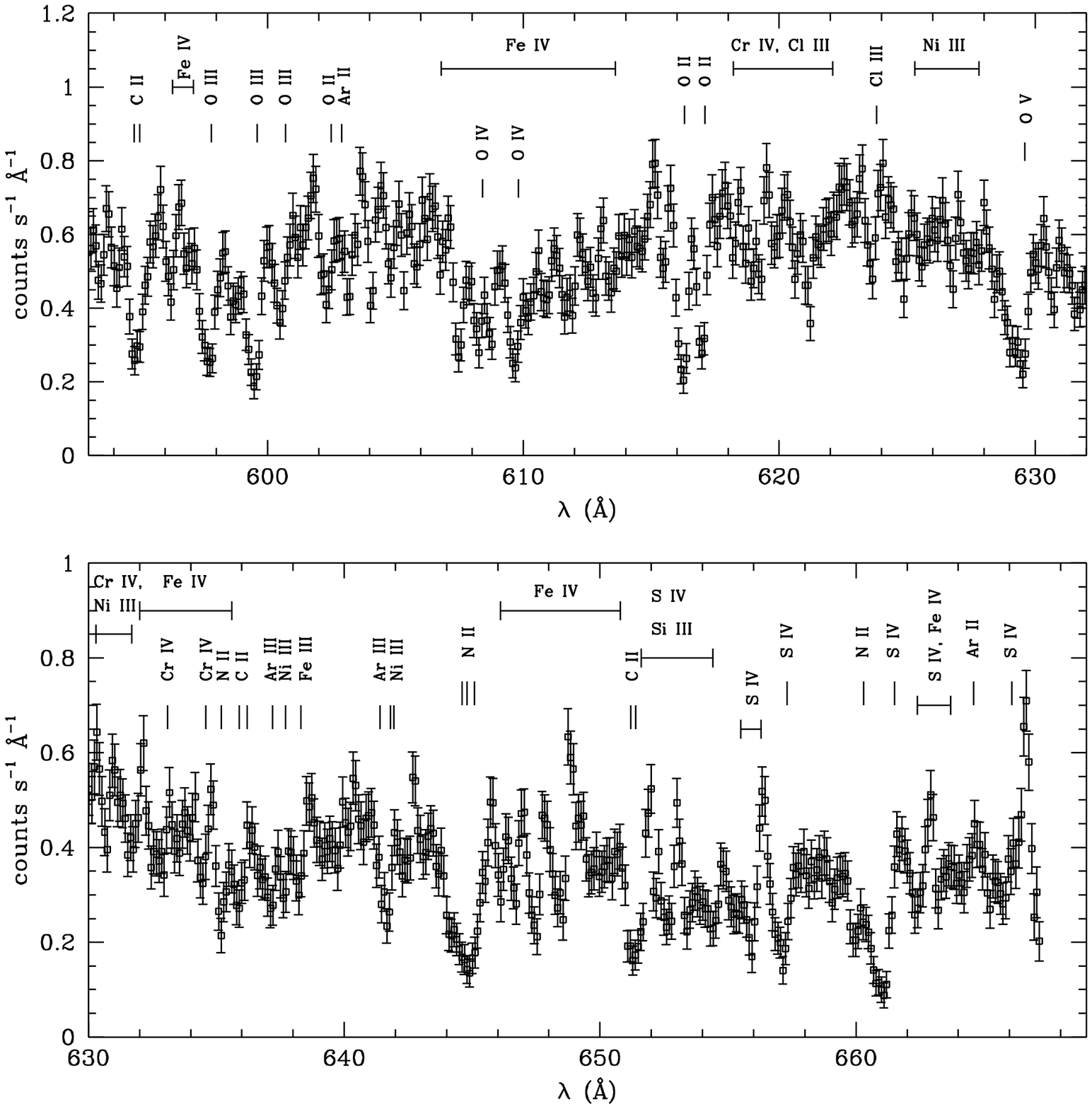}
\caption[]{continued}
\end{figure}

Despite the relatively high resolution of the data, the spectrum still
suffers from line-blending because of the very high density of features.
It is not clear if the continuum level is actually reached anywhere in
the \orfeus\/ spectrum, although it may be near 528 \AA, and possibly
a few other places. It should be noted that \ecma\ has a very low
projected rotational velocity of 35 km s$^{-1}$ \markcite{uf82}(Uesugi
\& Fukuda 1982), so that narrow photospheric lines are not resolved.
The strongest unambiguously identified features -- from \ion{C}{2},
\ion{N}{2}, \ion{O}{2}, and \ion{He}{1} -- are all predicted to be
among the strongest lines in the non-LTE 21,000 K model.  However, iron
seems to be more highly ionized (predominantly \ion{Fe}{4} rather than
\ion{Fe}{3}) than in the 21,000 K model, although the iron features are
often heavily blended and thus are difficult to identify individually.
The 21,000 K model makes relatively accurate predictions of the
absorption features even as it fails to match the continuum level by
nearly 2 orders of magnitude.  However, detailed modeling in which
individual line strengths are compared to the data will be necessary
before the validity of specific model characteristics can be
evaluated.  Fortunately, with the higher resolution available with the
\orfeus\/ data, these quantitative comparisons can now be made.

\begin{figure}
\vspace{-0.25in}
\plotone{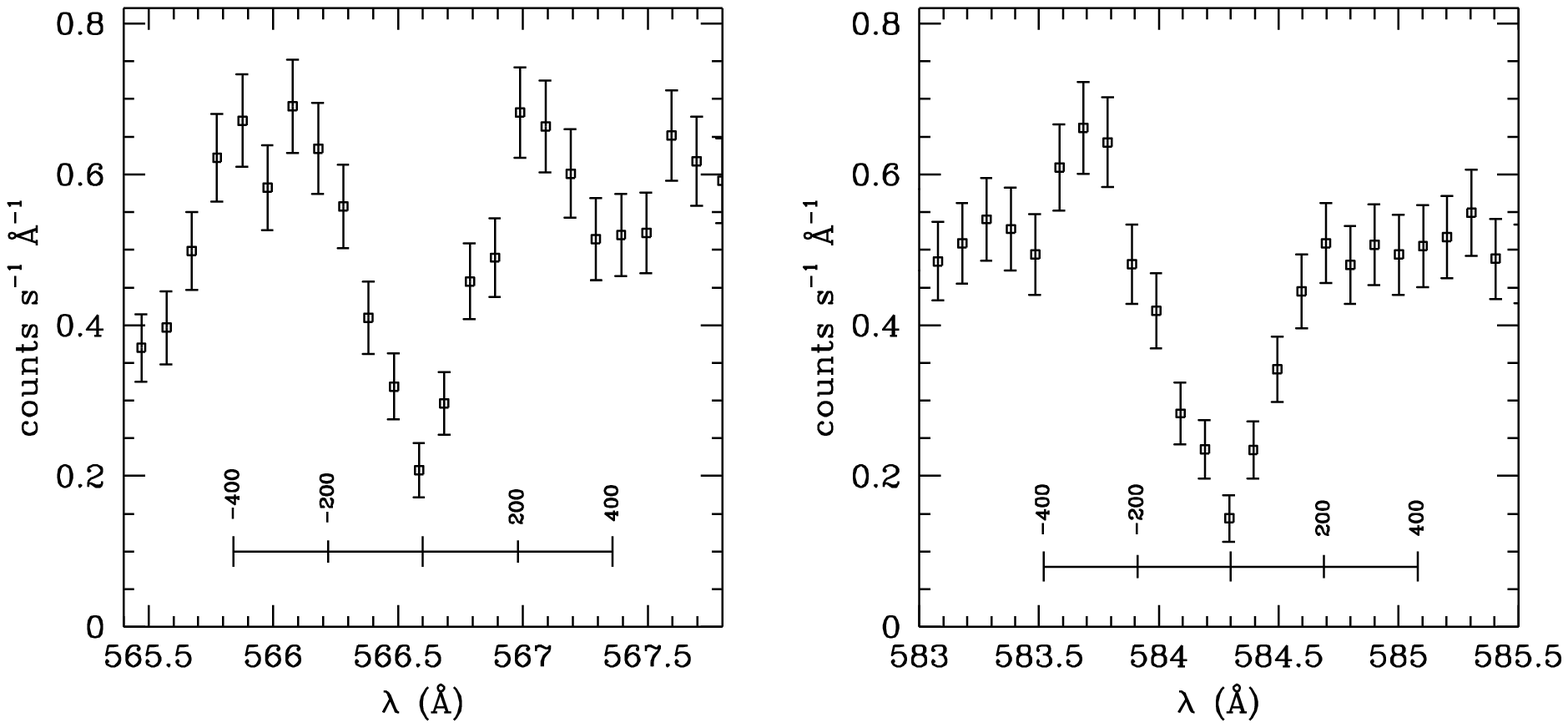}
\vspace{-3in}
\caption[]{Two of the best separated, unblended features in the
\orfeus\/ spectrum are shown for \ion{Si}{3} (left) and \ion{He}{1}
(right).  The velocity scale, in units of km s$^{-1}$, is indicated at
the bottom of each panel. These features are used to determine the
effective spectral resolution of the EUV spectrometer.}
\label{fig:res}
\end{figure}

To assess the resolution achieved in this data, we examined several
intrinsically narrow, relatively unblended lines and measured their
FWHMs.  The cleanest example of such a well-behaved feature is the
\ion{Si}{3} line at 566 \AA, shown in Figure \ref{fig:res}.  Its FWHM
is roughly 240 km s$^{-1}$, implying a resolution of $\lovdl \approx
1250$.  Other lines, such as the \ion{He}{1} line at 584 \AA, give
similar results.  This resolution is less than that achieved in the
far ultraviolet.

\placefigure{fig:res}

There are several possible explanations for the poorer than expected
resolution.  The off-axis position of the filtered aperture introduces
optical aberrations, but numerical ray traces indicate that these can
contribute only to a modest decline in the system resolution (from an
ideal of $\lovdl \approx 5000$ for on-axis spectra to about $\lovdl
\approx 4000$ for these off-axis pointings).  Simple Z-axis defocus
did not appear to be significant in the FUV \markcite{hur97}(see
Hurwitz \etal\ 1998), but these EUV spectra are dispersed by
diffraction grating B, an independent optic.  This opens the
possibility that Z-axis defocus may have been present.  However, all
four gratings were brought to a common focus during laboratory
calibration, and the Z-axis positions of the grating holders are
monitored electronically.  No significant Z-axis motions occurred
between the laboratory calibration and in-flight operations.  We note,
however, that the grating B spectrum was displaced from its preflight
position on the detector, indicating that the optic may have shifted
within its holder after laboratory calibration.  Such a shift could
introduce aberrations consistent with the observed resolution loss.
We also considered the possibility that the tin filter itself, which
is displaced from the focal plane by 2 mm in Z, broadens the effective
size of the point source.  However, the optical constants indicate
that the vast majority of the attenuated light should be absorbed (not
scattered), and the extreme thinness of the filter makes it unlikely
that the membrane affects the path of the transmitted rays to any
appreciable degree.

\begin{figure}
\plotone{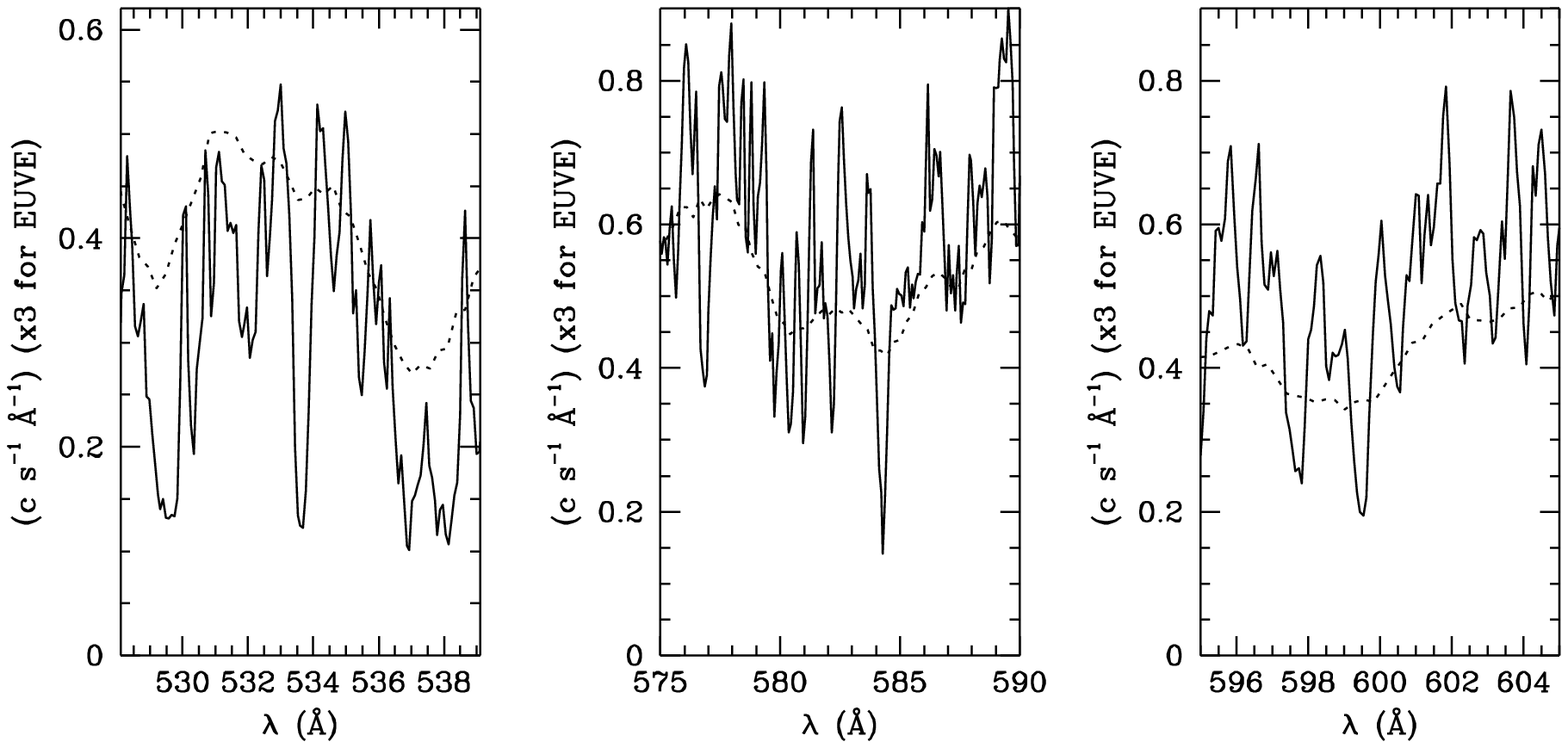}
\vspace{-3in}
\caption[]{We show several portions of the  \orfeus\/ spectrum (solid
  line) with the \euve\ data (dotted line) for \ecma\ overplotted in
  counts per unit wavelength interval.  The \euve\ spectrum is
  multiplied by 3 throughout in order to bring its level up to that
  seen in the \orfeus\/ data.}
\label{fig:euve}
\end{figure}

In Figure \ref{fig:euve}, we compare the \euve\ and \orfeus\/ spectra of
\ecma\/ in several wavelength regions.  The effect of the higher
resolution afforded by \orfeus\/ can be clearly seen in this figure,
with individual features in the \euve\ data now resolved into
multiple components in the \orfeus\/ data.  This increase in resolution
will be valuable for testing atmosphere models, which must reproduce
the rich spectral features in the EUV as well as the high continuum
levels if the mystery of the extraordinary continuum fluxes in very
early B giants is to be solved.  This figure also shows that the
\orfeus\/ EUV spectrometer is indeed about 3 times more sensitive
than the \euve\ long-wavelength spectrometer.  This general result
appears to hold over the entire common wavelength range of 520--665 \AA, although the relative sensitivity of \orfeus\/ increases
slightly at the long wavelength end of the range.

\placefigure{fig:euve}

\begin{figure}
\vspace{-1.75in}
\plotone{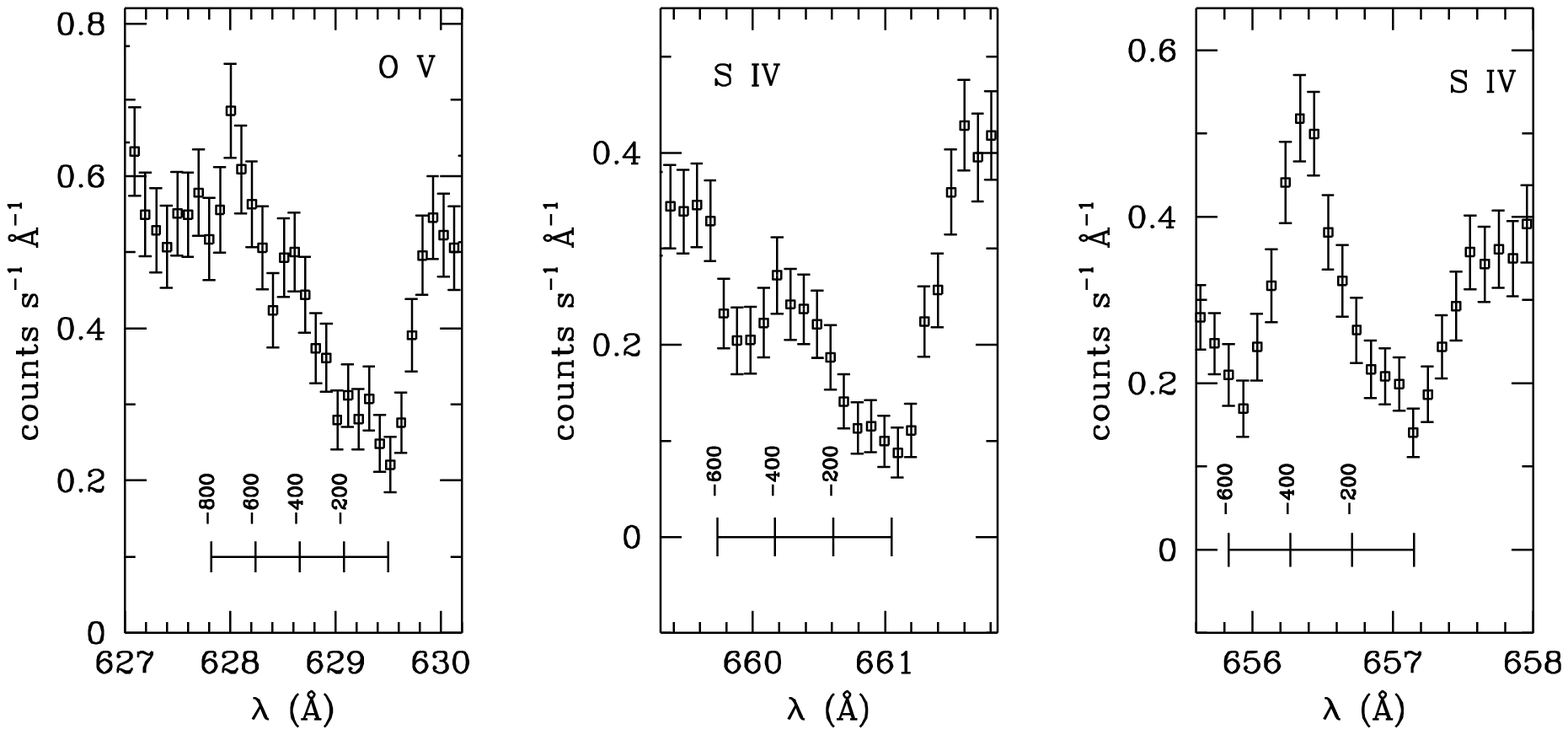}
\vspace{-3in}
\caption[]{Wind broadened oxygen and sulfur lines are shown, with a
velocity scale in units of km s$^{-1}$ indicated at the bottom of each
panel.  These display the classical asymmetric wind signature,
although the mass-loss rate of \ecma\ is low enough that no P Cygni
emission feature is seen.  Features that appear near $-500$ km
s$^{-1}$ are due to blends, not the wind.}
\label{fig:wind}
\end{figure}

With the improved resolution of \orfeus, wind features can be
resolved, whereas in the \euve\ spectra of \ecma\ the presence of
wind features could be inferred only from their high ionization stages
\markcite{cass95}(Cassinelli \etal\ 1995).  As Figure \ref{fig:wind}
shows, we can now see the distinctive wind-broadened morphology in the
\ion{O}{5} line at 630 \AA\ and the \ion{S}{4} doublet at 657, 661
\AA.  Although other \ion{S}{4} features are seen in \ecma, the
resonance lines at 657, 661 \AA\ are the most reliable indicators of
wind activity. It should be noted that some features of intermediate
ionization stages, such as the \ion{O}{4} lines near 554 and 610
\AA, are possibly asymmetric and slightly broader than the
intrinsically narrow lines used for the spectral resolution
determination.

\placefigure{fig:wind}

The blue-edge velocity of the \ion{O}{5} line lies somewhere between
500 and 800 km s$^{-1}$, depending on the true continuum level and the
degree of blending with other features.  This range of values is only
slightly below the theoretical terminal velocity, and is comparable to
the blue edge velocity seen in the \ion{C}{4} $\lambda\lambda$ 1548,
1551 doublet \markcite{sm76}(Snow \& Morton 1976).  The \ion{S}{4}
lines have blue edge velocities closer to 300 km s$^{-1}$.  This lower
velocity may be due to ionization gradients in the winds.  By
combining these EUV wind lines observed with \orfeus, including the
\ion{O}{4} features as well as \ion{O}{5} and \ion{S}{4}, with \iue,
{\it Hubble Space Telescope}, interstellar medium absorption profile
spectrometer (IMAPS), and \copernicus\ data, it should be possible to
place tight constraints on the mass-loss rate, terminal velocity, and
ionization balance of the wind of \ecma.

\section{Conclusions}

The performance of \orfeus\/ in the EUV was excellent, although the
spectral resolution is lower than that generally achieved in the FUV.
The effective area, including the tin filter transmission, is about 3
times that of \euve, as expected.  The data collected during the
flight of the telescope in 1996 pose a strong challenge to the models
currently being made to explain the large photospheric flux
of \ecma.  This is because equivalent widths of individual EUV lines
can now be determined.  The wind lines, primarily \ion{O}{5}, have now
been resolved and will provide important new constraints on the wind
ionization models of early B stars. Similar EUV spectroscopy of the
other known EUV-bright B giant, \bcma, would be very helpful in
determining the wind properties and the cause of the high EUV fluxes
in early B stars.

\clearpage

\acknowledgements

\centerline
{\bf Acknowledgments}

We wish to thank Chris Conselice for advice about
the line identification and wavelength solution, and Van Dixon and
Jean Dupuis for valuable discussions about the \orfeus\ instrument,
data reduction, and science issues.  This work was supported by NASA
grant NAG5-4761.

\end{document}